# An infinitely long flexible polymer chain in between two parallel plates

**Pramod Kumar Mishra**
Email: pkmishrabhu@gmail.com
Department of Physics, DSB Campus, Kumaun University,
Nainital,Uttarakhand (INDIA)

**Abstract**. We consider a fully directed self-avoiding walk model on a cubic lattice to mimic the conformations of an infinitely long confined flexible polymer chain; and the confinement condition is achieved by two parallel athermal plates. The confined polymer system is under good solvent condition and we revisit this problem to solve the real polymer's model for any length of chain and also for any separation in between the plates. The equilibrium statistics of the confined polymer chain is derived using an analytical calculations based on the generating function technique. The force of the confinement, the surface tension and the monomer density profile of confined chain is obtained. We propose that a method of calculations is suitable to understand thermodynamics of an arbitrary length confined polymer chain.

## 1. Introduction

The lattice model is widely used during past a few decades to understand the conformational statistics of a short polymer chain under confined geometries and also this model is used to understand statistics of a polymer chain in the bulk [1-3]. Therefore, there are variety of interesting results on the thermo-dynamical properties of a short and an infinitely long flexible polymer chain in the bulk and also under various confined geometries [4-7]. Such studies reveal wealth of information regarding scaling behaviour, phase transitions, local and universal properties of the polymer chains; and these reports gives us idea about the steric stabilisation of the polymer dispersions, colloidal solutions, thin films, surface coatings [3, 5, 7-9].

Though, there are couple of facts which are not well understood regarding an infinitely long and a short polymer chain for their three dimensional confined conformations; *e. g*., the scaling relations for the thermo-dynamical properties of a confined flexible chain is not well understood. Therefore, we have chosen a toy model (*i. e.* a directed walk model) of real polymer chain to understand the thermo-dynamical properties of an infinitely long and a short flexible polymer chain under proposed confined geometry. The confinement condition is established around the polymer chain using a pair of impenetrable flat plates; and the separation between the plates is measured in the unit of a monomer length. Thus, length of the polymer chain confinement is varied from one monomer length to the length of the polymer chain.

An exact number of the conformations of a short confined flexible homo-polymer chain is obtained using an underlying cubic lattice and a fully directed self avoiding walk [3] model is chosen to mimic the conformations of the confined flexible polymer chain of an *N* monomers, or in other words the polymer chain is lying in between two impenetrable parallel plates, as the confinement condition is schematically shown in the figure no. 1. A generating function technique is used to solve fully directed walk model analytically to study the conformational statistics of the confined an infinitely long as well as a short polymer chain.

The manuscript is organized in the following manner: a brief outline of fully directed walk model for present report is given in the section-2. The results of an analytical calculations are given in the section-3. The results are summarized and the conclusions are highlighted in the section-4.

## 2. The model and the method

A lattice model of fully directed self avoiding walk is widely used to understand the thermodynamics of an infinitely long polymer chain under various geometries [3, 10-13]. Since a directed walk model is solvable analytically and therefore we have an exact results about single chain statistics through such a toy (*i. e.* directed walk) model. It is well known that qualitative nature of the phase diagram for a directed walk model is same as its isotropic version of the problem [11-13]. It is assumed that first impenetrable plate is placed at $x=0$ and another impenetrable parallel plate is placed at $x=L$, where value of $L=1,2,3,....,\infty$; and, parameter $L$ is measured in the unit of the monomer length, and the confined polymer chain is schematically shown in the figure no. 1. A condition of $L \geq N$ corresponds to a polymer chain in the bulk.

It is well known that in the case of a fully directed walk model in three dimensions the walker is allowed to take steps only along $+x$, $+y$ and $+z$ directions [3] in between two parallel plates, and the maximum $L$ steps may be taken by the walker along $+x$ direction while walker can take any number of steps along remaining other two directions *i. e.* along $+y$ and $+z$. We have conformations of an $N$ monomers long polymer chain in between two parallel plates, where one end of the chain is grafted at the corner ($O$) of the lower plate. Thus, we have condition of the confinement provided $N>L$.

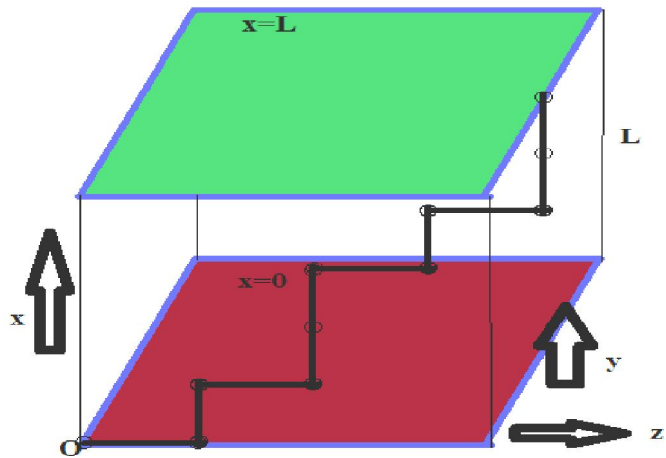

**Figure No. 1:** An $N$ monomers long flexible polymer chain is shown in this picture schematic, and the chain is confined by a pair of impenetrable parallel plates. The lower plate is located at $x=0$ and the upper plate is located at $x=L$; and one end of the polymer chain grafted at a point $O$ on the lower plate *i. e.* on the plate which is located at $x=0$.

A general expression for the grand canonical partition function of an infinitely long confined flexible chain is written as,

$$G(g,z) = \sum_{N=1}^{\infty} \sum_{All\ walks\ of\ N\ monomers} g^P z^{N-P} \quad (1)$$

A symbol *g* refers to the step fugacity of the walker parallel to the plane of the confining plates, while *z* is the step fugacity perpendicular to the plane of the confining plates. There are *P* monomers of the chain lying in the plane of the parallel plates and remaining (*N-P*) monomers are located perpendicular to the plane of the plates, for an *N* monomers long confined polymer chain.

## 3. The results

An equilibrium statistics of an infinitely long confined flexible polymer chain and also a short confined polymer chains may be obtained using lattice model [3]. We obtained exact results on the conformational statistics of a flexible polymer chain for its confinement using two parallel impenetrable plates; and analytical calculations are given below for a short chain and an infinitely long flexible polymer chains separately.

### A. An equilibrium statistics of an infinitely long confined flexible polymer chain

An exact expression of the grand canonical partition function for an infinitely long flexible polymer chain is obtained for different possible values of the plate separation and also for the bulk case; the partition function of the chain may be written as,

$$G(g,z) = \sum_{P=1}^{\infty}(2g)^P + \sum_{K=1}^{\infty} z^K \left\{1 + \sum_{M=1}^{\infty} \frac{(M+1)(M+2)\ldots(M+K)}{K!} * (2g)^M \right\} \quad (2)$$

We were able to recover an expression for the grand canonical partition function of the chain for the bulk case [11] by substituting *z=g* (when *L≥N* and *N→∞*) in the equation no. 2. The first term on the right hand side of equation no. 2 corresponds to the conformations of the chain lying on the lower plate (*i. e.* at *x=0*). A simple form of the expression for the grand canonical partition function for the bulk case (*i. e.* for an infinite separation between the parallel plates) may be written as,

$$G(g,z,L=\infty) = \frac{z+2g}{1-z-2g} \quad (3)$$

The partition function of an infinitely long polymer chain is obtained for the finite separation (*L*) in between the confining plates; and accordingly, we derived the thermodynamical properties of the confined chain. It is well known that the critical value of the step fugacity is 0.5 for the finite separation (*L*) in between parallel plates, and the critical value of the step fugacity is 0.33 for *L≥N* and *N→∞*.

### B. An equilibrium statistics of a short flexible polymer chain confined by a pair of parallel plates

We use canonical ensemble formalism to obtain an exact number ($C_N^L$) of a flexible polymer chain conformations and accordingly statistics for a case when the polymer chain is confined by a pair of impenetrable parallel plates; and an exact number of conformations is written as follows for a case when L<N,

$$C_N^L = 2^N + \sum_{L=1}^{N-1} 2^{N-L} \frac{N(N-1)(N-2)\ldots(N+1-L)}{L!} \quad (4)$$

We have number of conformations ($C_N^B$) of a chain for a case when this short flexible chain is in the bulk and number of the conformations for the bulk (*L≥N*) case is written as,

$$C_N^B = C_N^{L(=N-1)} + 1 = 3^N \quad (5)$$

An effect of the confinement is shown in the figure no. 2, where a fraction of the polymerized ($C_N^L$) and also a fraction of non polymerized ($1 - C_N^L$) short polymer chain conformations is shown for a set of values of the plates separation (*L*) and chain length *N*.

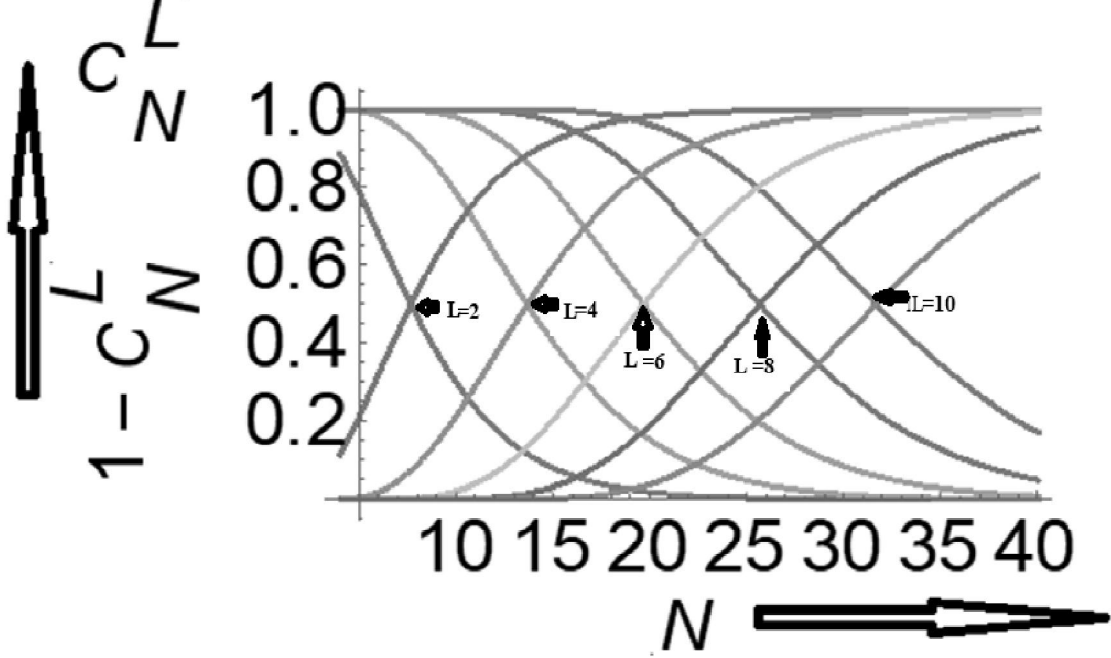

**Figure No. 2:** We have shown an average number of a flexible chain conformations of an *N* monomers long chain ($C_N^L$); and we have also shown in this figure an average number of the polymer chain conformations ($1-C_N^L$) which were not polymerized due to a pair of an impenetrable plates. The curves for $C_N^L$ and ($1-C_N^L$) intersects each other at 0.5 for a given value of *L* and *N*. As we increase the separation in-between the pair of plates ( *i. e.*, *L*=2, 4, 6, 8 and 10), the point of intersection moves to higher value of an *N*.

We have calculated the force of confinement ($f_N^L$) acting on a short polymer chain of *N* monomers due to parallel plates and perpendicular to the plane of the plates, where the separation in between the plate is *L* and free energy of the chain is written in a unit of the thermal energy as $E(=-k_B T Log[C_N^L])$; and thermal energy we have taken unity. Therefore, a graph between $f_N^L$ versus *L* and for an *N* is shown using figure no. 3. The force of the confinement is given by a following equation,

$$f_N^L \approx -Log[2] + \frac{\partial[Log\{\frac{\prod_{L=1}^{L}(N-L+1)}{L!}\}]}{\partial L} \tag{6}$$

Above equation is simplified to following relation to see that $f_N^L$ =-Log[2]-Log[$\frac{L}{N}$], thus for an infinitely long chain the force bear logarithmic singularity, provided *L*<*N*, *L*≥*1* and *N*→∞.

$$f_N^L \approx Log[\frac{N}{2L}] \tag{7}$$

The surface tension ($\sigma_N^L$) of a short and confined polymer chain solution may be obtained using following relation,

$$\sigma_N^L = \frac{\delta(E)}{\delta A} \tag{8}$$

Where $E(=-k_B T Log[C_N^L])$ is the free energy of a short polymer chain under confined geometry, and again we have taken value of the thermal energy ($k_B T$) equal to unity for the sake of mathematical simplicity. An areal change has value $[(N-L+1)^2-(N-L)^2]/2$ when the walker step one unit along *x* direction, and the plate separation varied in the unit of one monomer length. For a confined chain $L_{Max}=N-1$.

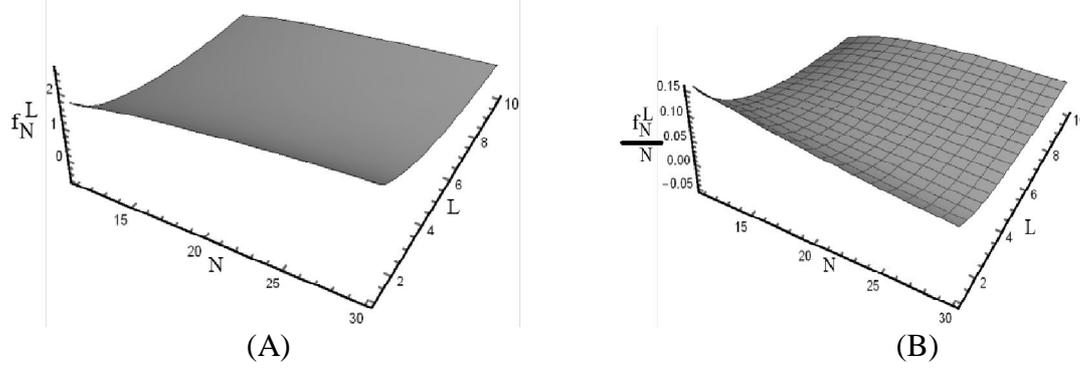

(A)                                               (B)

**Figure No. 3:** The force of the confinement which is acting perpendicular to the plane of the plates, and it's nature has been shown in this figure 3(A) for a short chain of length N monomers and we increase the separation in-between plates (L=2, 4, 6, 8 and 10) in a unit of monomer length. Figure 3(B) shows nature confining force per monomer of the confined chain.

$$\sigma_N^L = \frac{2\,Log[\frac{1}{2}]}{N-L} + \frac{2\,L\,Log[\frac{1}{N}]}{(N-L)^2} + \frac{Log[L!]}{(N-L)^2} \tag{9}$$

Nature of variation of the surface tension of a short polymer chain with confining plate separation is shown in figure no. 4.

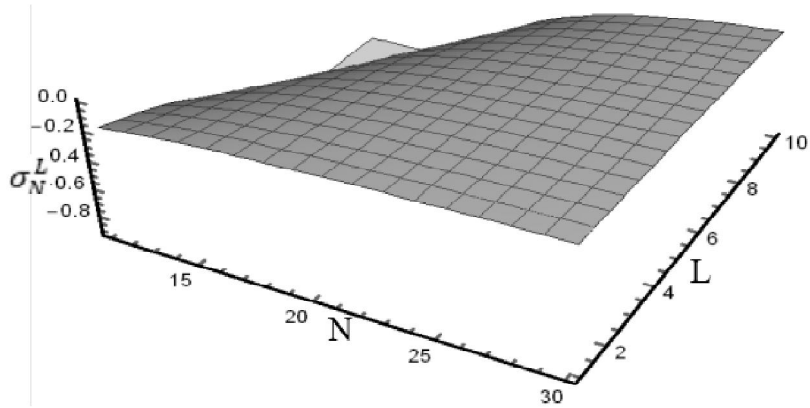

**Figure No. 4:** We have plotted the value of the surface tension ($\sigma_N^L$) of a short polymer chain (N=11, 12,...,30) confined in between a pair of parallel plates and the plates are separated by a distance L(=1,2,...,10).

We have calculated the monomer number density ($\rho_N^L$) profile plot for the confined flexible chain and we have an exact expression for the number density of an N monomers long chain; and the density is written as follows:

$$\rho_N^L = [2^{N+1-L}]\,\frac{N(N-1)\ldots(N-L+1)}{L!(N-L)} \tag{10}$$

The monomer number density profile is shown in the figure no. 5 for the confined condition of a short flexible polymer chain for given values of L. We have $\rho_N^L = \frac{2^{N+1}}{N}$ for L=0 and for non zero value of L and also for a confined chain (L<N), the monomer density profile is written as per equation no. 10.

## 4. The summary and the conclusions

A lattice model of the fully directed self-avoiding walk [3] is used to mimic the conformations of an infinitely long and also a short flexible polymer chain when the chain is confined by a pair of impenetrable parallel plates (as it shown schematically in figure no. 1). The confined regions in between the pair of plates leads different value of the step fugacity for the walker along and perpendicular to the plane of the plates. Therefore, along the plane of the plates we have one value of step fugacity ($g$) and in a direction perpendicular to the plane of the plates we have another value ($z$) of the step fugacity. We used a method of the generating function to solve the model analytically and we obtained a general expression for the grand canonical partition function of an infinitely long flexible polymer chain for any given value of plate's separation ($L$).

We have also obtained an exact expression of the canonical partition function for a flexible polymer chain of an $N$ monomers long chain for a given separation ($L$) in between the parallel plates. We calculated an exact percentage of the polymer conformations which were not polymerized due to confinement condition imposed on the chain by a pair of parallel impenetrable plates. We derived expressions for the force of the confinement, the surface tension and the monomer number density profile for a short chain of an $N$ monomers length and when this chain is confined by the pair plates for a plate separation $L$.

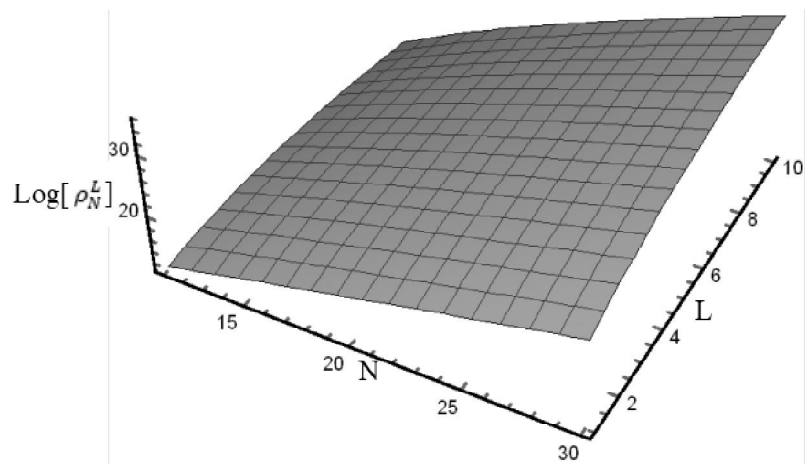

**Figure No. 5:** This figure shows the logarithmic value of the number density (*i. e.* number of the monomer per unit area of the confined region) of a flexible polymer chain. The length of polymer chain ($N$) varied from 11 to 30 and the separation ($L$) of the parallel plates is varied from 1 to 10.

We have plotted ($C_N^L$), the number of a confined flexible polymer chain conformations, of an $N$ monomers long polymer chain along with number of conformations (*i. e.* 1-$C_N^L$) which is suppressed due to the confinement, for different values of the plate separation (L); and we have shown it in figure no. 2. It is seen from this figure that for a given value of the plate separation ($L$); and as we increase the length of the chain, the percentage of the polymerized conformations decreases and accordingly percentage of suppressed conformations due to confinement increases. It is also found that the percentage of the polymerized conformations increases for a given separation in between the plates with an increase of chain length.

The force of confinement is a function of separations in between the plates and also this force is function of chain length. It is found from analytical calculations that the force of confinement decreases as we increase given value of the plate separation for a chain of an $N$ monomers. While the force of confinement increases with increase of chain length for a given value of the plate separation. We have shown the nature of confining force which is acting on a short polymer chain in figure no. 3 for a set of $N$ and $L$ values.

The surface energy per unit surface area (*i. e.* the surface tension) for a confined flexible chain is shown in the figure no. 4 and it is seen that the surface tension of the confined chain fluid increases for a given length of the chain as we increase length of confinement. The surface tension of a confined chain also increases for a given value of L as we increases number of monomers of the chain. Actual dependency of the surface tension on $N$ and $L$ is shown in the figure no. 4 and the mathematical form is given by an equation no. 9.

The monomer number per unit areal extension of the confined flexible chain is shown in the figure no. 5. It is seen from figure that the monomer density increases as the length of chain increases and L remain fixed, and also the monomer density increases as we increase L for a given length of the confined chain. The nature of free energy curve for a confined short chain is shown in figure no. 6 for the sake of completeness. It is seen from figure no. 6 that in the thermodynamic limit the free energy per monomer of confined flexible chain is -*Log[$g_c$(2D)]$^{-1}$*, where $g_c(2D)$=0.5.

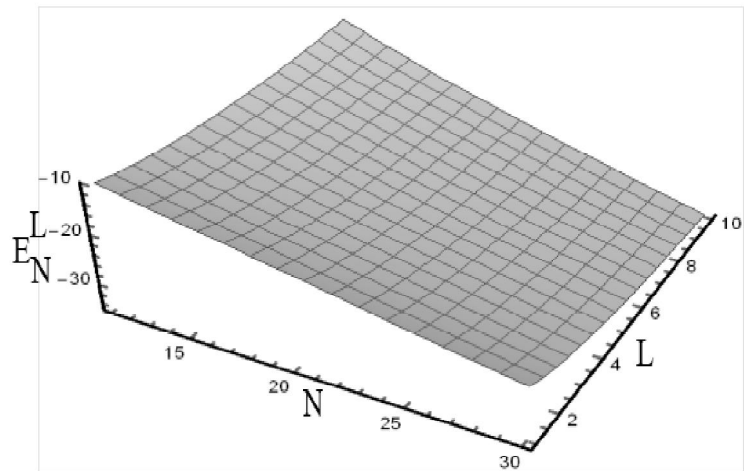

**Figure No. 6:** We have shown free energy of a confined flexible polymer chain for given values of *L* and *N*. The free energy is function of *N* and *L* and the free energy is approximated as $E \approx -NLog[2]+LLog[2]-LLog[N]+LLog[L]-L$. Thermal energy is set to unity for the sake of mathematical simplicity.

A method of calculations reported in this manuscript may be easily extended to calculate the thermodynamics of an infinitely long as well as a short polymer chain confined to a length L for other versions of the directed walk models on other possible lattices.